\documentclass[10pt,conference]{IEEEtran}

\usepackage{cite}
\usepackage{amsmath,amssymb,amsfonts}
\usepackage{algorithmic}
\usepackage{graphicx}
\usepackage{textcomp}
\usepackage{xcolor}
\usepackage{braket}
\usepackage{hyperref}
\usepackage[qm]{qcircuit}
\usepackage{stfloats}
\newcommand{\mycomment}[1]{}

\def\BibTeX{{\rm B\kern-.05em{\sc i\kern-.025em b}\kern-.08em
    T\kern-.1667em\lower.7ex\hbox{E}\kern-.125emX}}
\begin{document}

\title{Feedback Connections in Quantum Reservoir Computing with Mid-Circuit Measurements\\}

\author{\IEEEauthorblockN{Jakob Murauer\IEEEauthorrefmark{1}, Rajiv Krishnakumar\IEEEauthorrefmark{2}\IEEEauthorrefmark{3}, Sabine Tornow\IEEEauthorrefmark{1} and Michaela Geierhos\IEEEauthorrefmark{1}}
\IEEEauthorblockA{\IEEEauthorrefmark{1}Research Institute CODE, University of the Bundeswehr Munich, 85579 Neubiberg, Germany}
\IEEEauthorblockA{\IEEEauthorrefmark{2}QuantumBasel, 4144 Arlesheim, Switzerland}
\IEEEauthorblockA{\IEEEauthorrefmark{3}Center for Quantum Computing and Quantum Coherence (QC2), University of Basel, 4001 Basel, Switzerland}
}

\maketitle
\begin{abstract}
 Existing approaches to quantum reservoir computing can be  broadly categorized into restart-based and continuous protocols. Restart-based methods require reinitializing the quantum circuit for each time step, while continuous protocols use mid-circuit measurements to enable uninterrupted information processing. A gap exists between these two paradigms: while restart-based methods naturally have high execution times due to the need for circuit reinitialization, they can employ novel feedback connections to enhance performance. In contrast, continuous methods have significantly faster execution times but typically lack such feedback mechanisms. In this work, we investigate a novel quantum reservoir computing scheme that integrates feedback connections, which can operate within the coherence time of a qubit. We demonstrate our architecture using a minimal example and evaluate memory capacity and predictive capabilities. We show that the correlation coefficient for the short-term memory task on past inputs is nonzero, indicating that feedback connections can effectively operate during continuous processing to allow the model to remember past inputs.
\end{abstract}


\section{Introduction}
\label{sec:intro}

Reservoir computing is a machine learning approach for predicting time series data \cite{best_practice, class_reser, class_reser_2, class_reser_3}. A reservoir computer transforms an input sequence into a higher dimensional feature space using a fixed but random non-linear system called a reservoir. A simple linear regression model is then trained to read the state of the reservoir and map the result to a desired output. Since the training is done only in the last stage, while the internal dynamics are fixed, the optimization is very efficient. Quantum Reservoir Computing (QRC) is the quantum analog of classical reservoir computing, where a quantum system is used as a reservoir and its complex dynamics are used to efficiently process temporal data \cite{quant_reser_1, quant_reser_2, quant_reser_3, quant_reserv_4, feedbackdriven, weak_meas, rewind_protocol}. This is seen as a promising approach because quantum systems offer a naturally high-dimensional Hilbert space and complex non-linear dynamics. These properties are thought to enhance the performance of reservoir computing \cite{quantum_dyn, feedbackdriven, boosting}. 

\begin{figure*}[!ht]
    \centering
    \includegraphics[width=\textwidth]{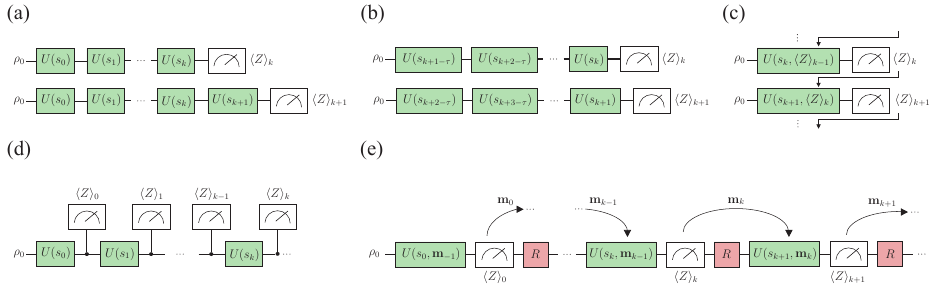}
    \caption{\textbf{QRC protocols:} Each architecture is designed to process an input sequence ${s_k}$. The systems are initialized in the state $\rho_0$. A unitary operator $U(\cdot)$, constructed from parameterized Pauli rotations, governs the state evolution. Each $U(\cdot)$ represents the processing of one time step (cycle). \textbf{(a)} Graphical representation of a conventional restart-based QRC scheme. For each timestep, the architecture must be restarted and run from the beginning. This results in a quadratic runtime. \textbf{(b)} A rewind protocol that does not restart from the beginning, but only processes the last $\tau$ time steps. \textbf{(c)} A feedback-controlled QRC protocol. Only one timestep is processed at a time, but the expectation value from the previous trial is stored in a buffer. \textbf{(d)} A mid-circuit measurement architecture that continuously monitors the quantum state. No restart is required. This architecture can be combined with weak measurements.  \textbf{(e)} A feedback-driven QRC architecture with mid-circuit measurements. Feedback is incorporated by feeding the measurement strings $\textbf{m}_k$ back into the reservoir, a process that can be done at runtime. The expectation values $\langle Z \rangle_k$ are then computed in the post-processing step after all the circuits have been executed.}
    \label{fig:All_QRC_Regimes}
\end{figure*}

Several architectures have been proposed, broadly categorized into restart-based and continuous methods. Conventional restart-based QRC schemes generate the output time series by averaging the measurement results over all qubits at each time step. This requires repeatedly restarting the system for each time step to obtain the complete time series, see Fig.~\ref{fig:All_QRC_Regimes} (a). To avoid running the whole process from the beginning, methods known as rewind protocols have been proposed \cite{weak_meas}. Here, the reservoir for cycle $k$ is initiated from cycle $k+1-\tau$, thus reducing the computation to the last $\tau$ cycles compared to the first cycle in the conventional approach (Fig. 
~\ref{fig:All_QRC_Regimes} (b)). This method is motivated by the fact that reservoirs have a fading memory property \cite{Maass2002-wv}, which means that over time the reservoir loses information from distant time steps, making them insignificant for more recent cycles. On the other hand, continuous QRC methods use mid-circuit measurements to continuously monitor the dynamics of the reservoir, Fig.~\ref{fig:All_QRC_Regimes}(d), which immensely improves the overall execution time on quantum hardware \cite{mid_circuit_measurements}. This method has also been combined with weak measurements to mitigate the erasure of past inputs, as they extract limited information without completely collapsing the quantum state. However, this comes at a cost, as the readout from the reservoir is now incomplete \cite{weak_meas}.

A novel approach introduces a feedback link by feeding back expectation values into the reservoir \cite{feedbackdriven}. This approach still restarts at each time step, but only needs to run for one time step at a time. While this method effectively incorporates memories of previous inputs into the reservoir, it still has a drawback, namely execution time on quantum hardware. As shown in Fig.~\ref{fig:All_QRC_Regimes} (c), for each time step the expectation value has to be estimated with a certain number of shots. Recently, a preprint has been published describing a modification of this protocol using weak measurements, where part of the reservoir state is carried over between time steps \cite{https://doi.org/10.48550/arxiv.2503.17939}. However, this method still requires many measurements to be taken between time steps in order to compute expectation values before feeding them into the next time step.

In this paper, we introduce a novel feedback-driven quantum reservoir computing approach with mid-circuit measurements and classical feedforward operations during the coherence time of a qubit. In the restart-based feedback protocol, as shown in Fig.~\ref{fig:All_QRC_Regimes} (c), each cycle requires $n$ shots. Thus, for an input sequence of training length $t$ tested over $k$ different reservoir realizations, the total number of shots scales with $t\times k\times n$. As $t$, $k$, or preferably $n$ grow, the cumulative cost of running these circuits becomes large. This motivates our exploration of feedback circuits operating within the coherence time of a qubit using novel classical feedforward methods. To isolate and rigorously quantify the effect of mid-circuit feedback, we perform a qubit reset after each time step. This ensures that no extraneous information (other than the intended feedback) propagates forward, allowing us to study the feedback coupling itself in a controlled manner. One can omit these resets to explore more complex scenarios in which correlations persist over time, but here our primary goal is to study feedback connections in isolation and thereby clarify their implications. For implementation, IBM superconducting systems have in particular demonstrated classical real-time feedforward operations \cite{dyn_circ_1}, providing an opportunity to reduce the overall shot overhead for certain problems. We use the measurement strings obtained from a mid-circuit measurement as feedback for the next cycle. We can tune the feedback strength and the input strength through hyperparameters. We evaluate the memory performance by studying the short-term memory task and the predictive capabilities by predicting the Mackey-Glass system \cite{Macke-Glass} and the expectation value of a spin in a chaotic Ising chain.

\section{Methods}
In the following, we provide an overview of our approach in three parts: first we describe the proposed mid-circuit QRC architecture, then we present the training procedure, and finally we explain the testing routine.

\subsection{Architecture}\label{architecture}
\begin{figure}[!hb]
    \centering
    \includegraphics[width=0.45\textwidth]{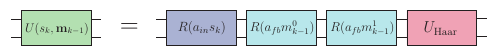}
    \caption{Minimal two-qubit unitary circuit. The circuit processes the input $s_k$ via a rotation $R(a_{\mathrm{in}}\,s_k)$ and then incorporates feedback via rotations $R(a_{\mathrm{fb}}\,m_{k-1}^0)$ and $R(a_{\mathrm{fb}}\,m_{k-1}^1)$, where $\{m_{k-1}^0, m_{k-1}^1\}$ are the previous measurement results. The structure of $R(\cdot)$ is explained in Eq. \ref{eq1}. A final Haar-random unitary $U_{\text{Haar}}$ is applied at the end.}
    \label{fig:Circuit}
\end{figure}

We propose a mid-circuit QRC architecture that incorporates feedback links, as shown in Fig.~\ref{fig:All_QRC_Regimes} (e). We start with a quantum system of $N$ qubits, initialized in the state $\rho_0^N$ which is chosen to be $\ket{0}\bra{0} ^{\otimes N}$ throughout this work unless otherwise mentioned. Given an input sequence ${s_k}$, the system evolution is governed by the unitary operator $U(s_k, \textbf{m}_{k-1})$, where $s_k$ is the current time step of the input sequence and $\textbf{m}_{k-1} \in \{-1,1\}^N$ are previous measurement results of the cycle $s_{k-1}$. The input sequence is inserted into a two-qubit gate $R(a_{in}s_k)$, while each previous measurement result $m_{k-1}^j$ is inserted into another two-qubit gate $R(a_{fb}m_i^j)$. The structure of the $R$ gate is defined as follows:
\begin{equation}\label{eq1}
    R_{i,j}(\theta) = CX_{ij}RZ_j(\theta)CX_{ij}RX_j(\theta)RX_i(\theta)
\end{equation}
Here $\theta$ represents an angle parameter that determines the rotation applied in the two-qubit gate. Specifically, $\theta$ can take values from the set $\{a_{in}s_k, a_{fb}m_{k-1}^j\}$, where $a_{in}$ and $a_{fb}$ are scaling parameters that control the input and feedback strength respectively. The structure of using $R$ gates along with the scaling parameters is commonly used in QRC models due to its compromise between being hardware efficient and generating substantial entanglement \cite{feedbackdriven, mid_circuit_measurements, naturalQRC}.
The full unitary $U(s_k, \textbf{m}_{k-1})$ consists of an input $R$ gate, $N$ feedback $R$ gates, and a Haar-random unitary. The idea is illustrated for the case of $2$ qubits in Fig.~\ref{fig:Circuit}. 
For the first time step $s_0$, a random vector $\textbf{m}_{-1}$ uniformly distributed from $\{-1,1\}^N$ is selected for time step $0$. After unitary evolution, the system is measured projectively in the computational basis. We note that further research can also employ weak measurements, as seen in other QRC research \cite{weak_meas, mid_circuit_measurements}. We reset our system to $\rho_0^N$ and feed the measurement string forward into the next unitary.

\subsection{Training}
\label{sec:training}
The task of QRC is to build a regression model that accurately approximates a target sequence \(\{\hat{y}_k\}\) for a given input sequence \(\{s_k\}\). In typical QRC mechanisms, we run the model for $l_w+l_{tr}+l_{ts}$ cycles, where $l_w$, $l_{tr}$, and $l_{ts}$ correspond to the lengths of the three main stages: washout, training, and testing respectively. As discussed in Section~\ref{architecture}, we start the first cycle with input \(s_0\) and a randomly selected feedback vector. To remove the effects of this initialization, we discard the first \(l_w\) measurement results (the \emph{washout} phase). A regression model is then built using the next \(l_{\mathrm{tr}}\) steps (the \emph{training} phase).
The linear regression model is built based on the reservoir outputs of the training phase. Specifically, we build the matrix $X_{tr} \in \mathcal{R}^{l_{tr} \times (N+1)}$, which is defined as:
\begin{equation}
    X_{tr} = \begin{bmatrix}
\langle Z \rangle_{lw+1}^1 & \langle Z \rangle_{lw+1}^2 & \cdots & \langle Z \rangle_{lw+1}^N & 1 \\
\langle Z \rangle_{lw+2}^1 & \langle Z \rangle_{lw+2}^2 & \cdots & \langle Z \rangle_{lw+2}^N & 1 \\
 &        & \vdots  &  \\
\langle Z \rangle_{lw+ltr}^1 & \langle Z \rangle_{lw+ltr}^2 & \cdots & \langle Z \rangle_{lw+ltr}^N & 1
\end{bmatrix}
\end{equation}where $\langle Z \rangle_k^n$ is the expectation value of the Pauli $Z$ operator on the $n^{th}$ qubit at the $k^{th}$ time step.

We now create a linear model $w_{opt}$ such that $y_{tr} = X_{tr} w_{opt}$ closely resembles the target sequence $\hat{y}_{tr}$. We use the Moore-Penrose inverse \cite{moore1920reciprocal,penrose1955generalized}, also called the pseudoinverse, to compute $w_{opt}$:
\begin{equation}
    w_{opt} = (X_{tr}^T X_{tr})^{-1} X_{tr}^T\hspace{1 mm} \hat{y}_{tr}
\end{equation}

\subsection{Testing}
\label{sec:testing}
We now test the trained model by using it to obtain a prediction vector of length  $l_{ts}$.
We create a matrix $X_{ts} \in \mathcal{R}^{l_{ts} \times (N+1)}$, which is defined as: 
\begin{equation}
    X_{ts} = \begin{bmatrix}
\langle Z \rangle_{lw+ltr+1}^1  & \cdots & \langle Z \rangle_{lw+ltr+1}^N & 1 \\
\langle Z \rangle_{lw+ltr+2}^1  & \cdots & \langle Z \rangle_{lw+ltr+2}^N & 1 \\
 &        & \vdots  &  \\
\langle Z \rangle_{lw+ltr+lts}^1 & \cdots & \langle Z \rangle_{lw+ltr+lts}^N & 1
\end{bmatrix}
\end{equation}
The prediction vector in terms of $w_{opt}$ is then given as:
\begin{equation}
    y_{pred} = X_{ts}w_{opt}
\end{equation}
The resulting prediction $y_{pred}$ is then compared to the target vector $\hat{y}_{ts}$ to evaluate the prediction performance of the model.

\section{Experiments}
In this section, we investigate the performance of our proposed reservoir computing model through three key experiments: measuring its short-term memory capacity, evaluating its predictive capabilities, and analyzing the echo state property.
\subsection{Short-term memory capacity}
\label{sec:stmc}
To characterize the performance of remembering past inputs we use the short-term memory capacity \cite{stm}. Consider an input sequence $\{s_k\}$, where each $s_k$ is drawn 
independently from a uniform distribution on $[0,1]$. 
The goal is to retrieve values of the sequence, formulated as
\(\hat{y}_k = s_{k+\tau}\), where \(\tau\) is zero or negative. We use the coefficient of determination, given by:
\begin{equation}
    R^2_\tau = \frac{\text{cov}^2(\hat{y}_{ts}, y_{pred})}{\sigma^2(\hat{y}_{ts}) \sigma^2(y_{pred})},
\end{equation}
as a figure of merit. When $R^2_\tau$ is close to 1, the model's predictions closely match the actual targets, indicating minimal error. In contrast, when $R^2_\tau$ is close to 0, the predicted outputs show substantial deviation from the actual targets.
Short-term memory capacity is then measured by summing $R^2$ for different time steps:
\begin{equation}
    C_{\Sigma} = \sum_\tau R^2_\tau
\end{equation}

\subsection{Predictive Capabilities}
\label{sec:pred-cap}
To evaluate the predictive capabilities, we test the model on a classical task and a quantum time series. For the study of classical dynamical systems, we choose the Mackey-Glass time series, a well-known benchmark that has been extensively used in the context of reservoir computing \cite{ESN-Mackey, feedbackdriven, quant_reser_1, ESN_Mackey_2}. The delay differential equation is defined as \cite{Macke-Glass}:
\begin{equation}
    \frac{dP(t)}{dt} = \frac{\beta_0 \theta^nP(t-\tau)}{\theta^n + P(t-\tau)^n} - \gamma P(t)
\end{equation}
We set $\beta_0 = 0.2$, $\theta = 1$, $n=10$, $\tau = 17$ and $\gamma = 0.1$. To generate the time series, we numerically integrate the delay differential equation using a forward Euler scheme with a time step of $\Delta t = 1$. The initial values are set to a constant for the duration of the delay, and the trajectory is normalized to the interval $[0,1]$.

The quantum time series is constructed from the expectation value of the $\sigma^z$ operator of the middle spin in a one-dimensional Ising chain governed by the Hamiltonian:  
\begin{equation}
    H = J\sum_{i}\sigma^z_i \sigma^z_{i+1} + h_x\sum_{i}\sigma^x_i + h_z\sum_i \sigma_i^z
\end{equation}  

We simulate the system using this Hamiltonian with parameters set to \( N = 5 \), \( J = 1 \), \( h_x = 1.05 \), and \( h_z = -0.5 \). This ensures that the quantum Ising model becomes non-integrable and exhibits chaotic behavior \cite{PhysRevLett.64.2215,chaotic}. At each time step, the expectation value \( \langle Z \rangle_t^2 \) of the central spin is measured, generating the time series. The time evolution is discretized with a time step of \( \Delta t = 0.005 \).  

For all prediction tasks, the goal is to predict steps in the future, similarly formulated with the target sequence as \(\hat{y}_k = s_{k+\tau}\), where \(\tau\) is positive. We evaluate the performance of the models with the norm mean squared error, which is defined as:
\begin{equation}
    \text{NMSE}(\hat{y}_{ts}, y_{pred}) 
    = \frac{\sum_{i=1}^N \bigl(\hat{y}_{ts,i} - y_{pred,i}\bigr)^2}
           {\sum_{i=1}^N \bigl(\hat{y}_{ts,i}\bigr)^2}
\end{equation}
The NMSE measures the size of the error relative to the scale (magnitude) of the true data. A small error indicates good model performance, while large values indicate a large deviation from the target sequence.

\subsection{Analysis of Echo State Property}
\label{sec:esp-method}
As mentioned in Section~\ref{sec:intro}, one of the important aspects of a reservoir computing model is the fading memory property. In the classical setting of echo state network (ESN) reservoir computing models, this can be implemented by ensuring that the ESN contains the echo state property (ESP) \cite{jaeger2001echo}, which ensures that the internal state of the model becomes asymptotically dependent only on the input history, regardless of its initial conditions. However, in quantum reservoir models, the fading memory property is usually generated by including some (preferably controlled) dissipation in the model \cite{PhysRevApplied.8.024030,Chen2019-nt}. Recently, there have been proposals of modified versions of the ESP that are specific to quantum reservoirs \cite{PhysRevE.110.024207}. In this paper, we describe and investigate a version of the ESP that applies across any reservoir computing model.

As a first step in investigating the role of the ESP in quantum reservoir computing models, we perform an analysis of the asymptotic dependence of the reservoir state of several reservoir computing models (including the one proposed in this paper) on their initial internal states. To do this, we study how the internal state of the reservoir evolves over time under different random initializations. We can define the internal state of the reservoir the same way as is defined in traditional reservoir computing models, i.e., $x_t \in \mathcal{R}^M$ at time $t$ to be
\begin{equation}
    x_t = F(x_{t-1},s_t)
\end{equation}
where $M$ is the reservoir dimension, $F$ is the fixed reservoir function, and $s_t$ is the time series at time $t$ \cite{jaeger2001echo,Maass2002-wv}.\footnote{In our architecture, $x_t$ = $[ \langle Z \rangle ^1_t , \langle Z \rangle ^2_t, \cdots, \langle Z \rangle ^N_t]$.}.
For each model, we use input sequences drawn from a uniform distribution on $[0,1]$ (as described in Section~\ref{sec:stmc}) over 5 random initializations of the model. For each run, we extract the first component of the internal state of the reservoir, $x_t^1$, at time $t$ and track its evolution over time across the different runs. To quantify the asymptotic behavior of these internal states, we compute the average difference of this component across runs. We perform this analysis for a classical minimal ESN model \cite{Lukoeviius2012}, a feedback-driven QRC protocol \cite{feedbackdriven}, a QRC scheme with a mid-circuit measurements architecture \cite{mid_circuit_measurements}, and the model proposed in this paper.

\section{Results and discussion}
In this section, we report on three key topics: simulation results of our proposed model, a hardware implementation on a QPU, and a simulation-based echo state property analysis for several reservoir computing models.
\subsection{Memory capacity and predictive performance}
We obtain all predictive and memory results by simulating a minimal instance of the proposed architecture with a system size of \(N=2\). We use a shot-based approach, where each shot represents a single experiment on the quantum hardware, and individual measurement results are sampled directly. This approach mimics a noiseless version of what is done on real hardware, and allows us to take into account the effect of shot noise which plays an important role in the performance of quantum reservoir models \cite{Sannia2024}. We also note that since we use a density matrix formalism, analytically considering all possible measurement trajectories under the conditional feedforward mechanism would require exponential overhead and is therefore not feasible. For further computational efficiency, we choose a washout length \(l_w = 25\), a training length \(l_{\mathrm{tr}} = 100\), and a test length \(l_{\mathrm{ts}} = 100\).

\begin{figure}[!ht] 
    \centering
    \includegraphics[width=0.48\textwidth]{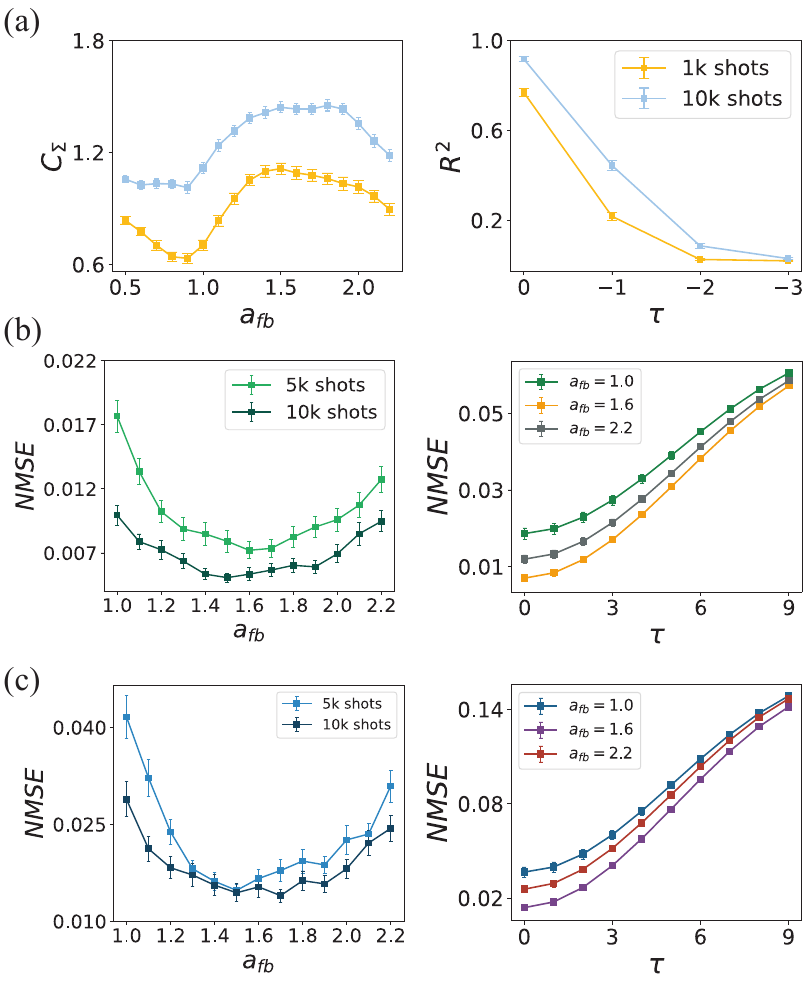}
    \caption{\textbf{Evaluation of short term memory and predictive capabilities:} All values are averaged over 128 Haar-random unitaries. Since there is no single `privileged' choice of unitary, our primary interest is in the average performance of the protocol over these random instances. Therefore we choose to indicate one standard deviation of the mean with the error bars to show how much this mean would vary if we resampled the Haar measure many times. \textbf{(a)} The left panel shows the short-term memory (STM) capacity for several different values of \(a_{fb}\), with \(\tau\) ranging from \(0\) to \(-3\). The right panel provides a more detailed view for \(a_{fb} = 1.3\), plotting all values of \(\tau\). \textbf{(b)} Here, we plot the NMSE for the quantum 1D Ising chain, showing the prediction capabilities for \(\tau = 1\) over different \(a_{fb}\) values. The right panel shows the prediction capabilities for different \(\tau\) values for selected $a_{fb}$ values. \textbf{(c)} shows the same analysis for the Mackey-Glass task. \(5k\) shots were used for the right panels in  both \textbf{(b)} and \textbf{(c)}.}
    \label{fig:results}
\end{figure}

In Fig.~\ref{fig:results} (a) on the right, we plot the short-term memory correlation coefficient for four different delay values for a fixed $a_{fb} = 1.3$.
Even with a limited number of shots, the minimal example shows nonzero correlation coefficients for \(\tau = -1\), indicating that the basic feedback connections during continuous processing play a beneficial role for memory. Notably, the feedback connections were evaluated using a reset operation after each measurement, ensuring that any information carried forward is solely from the proposed mechanism. Furthermore, an upward shift is observed for an increasing number of shots, suggesting that performance improves with additional measurement statistics.
On the left in Fig.~\ref{fig:results} (a) we show the effect of varying the feedback strength on the memory capacity \(C_{\Sigma}\) for different numbers of measurements. Again, we observe an upward shift in performance with increasing number of measurements. In particular, we believe that careful tuning of both feedback and input strengths is crucial for robust performance, as is the right choice of Haar matrix. As noted in \cite{feedbackdriven}, if the angles in the feedback and input \(R\) gates exceed \(\pm 2\pi\), the reservoir cannot reliably process the feedback connections and the memory capacity disappears. From the analysis in Fig.~\ref{fig:results} (a) we can see that there is an increase in memory capacity for $a_{fb}$ values from $1.0$ to about $2.0$. Interestingly, this trend is also reflected in the prediction performance -- lower NMSE values are observed for the same $a_{fb}$ range, as shown in Fig.~\ref{fig:results} (b) and (c). This again confirms the importance of memory capacity for predictive tasks. We can report an overall minimum average NMSE of $5.07 \times 10^{-3}$ on the 5-qubit Ising chain. The model performs slightly worse on the classical dynamical Mackey-Glass sequence with a minimum average NMSE of $1.4 \times 10^{-2}$. We also note that the model is robust to some amount of depolarizing noise (more details in Appendix~\ref{sec:noise-test}).

\begin{figure*}[ht]
    \centering
    \includegraphics[width=\textwidth]{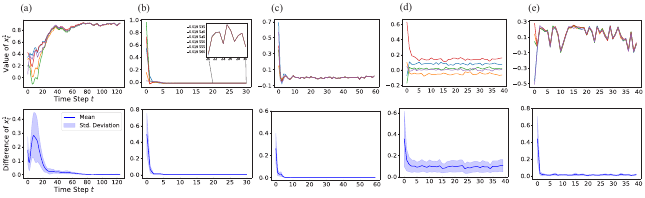}
    \caption{\textbf{Echo state property for different architectures}: The top figures show the first component of the reservoir state over time for five randomly initialized runs, while the bottom figures show the absolute difference between these five reservoir states. \textbf{(a)} Classical ESN model. \textbf{(b)} Feedback-driven QRC regime \cite{feedbackdriven}. \textbf{(c)} Mid-circuit measurements QRC \cite{mid_circuit_measurements}. \textbf{(d)} Proposed feedback-based QRC with mid-circuit measurements and post-measurement resets. \textbf{(e)} Proposed feedback-based QRC with mid-circuit measurements and continuous operation (no resets).}
    \label{fig:Convergence}
\end{figure*}
\subsection{QPU results}
We now present experiments with our architecture on IBM hardware, specifically utilizing the dynamic circuit capabilities of ibm\_marrakesh. Due to hardware constraints on the number of resets and conditional operations, we reduced the washout, training, and testing lengths to 5, 25, and 20, respectively. Additionally, time and hardware limitations made it infeasible to average over 128 Haar-random unitaries, so we selected a single fixed unitary. Below, we briefly describe how we generate a Haar-random unitary for the IBM device. A more detailed description of this algorithm is found in \cite{OliviaDiMatteo2021} and \cite{Haar_random}:
\begin{enumerate}
    \item Generate a $2^N \times 2^N$ matrix $A$ with complex entries $z_i = a_i + b_ii$, where all $a_i, b_i$ are sampled from the \textit{Ginibre} ensemble
    \item Compute the QR decomposition $A = QR$
    \item Compute the diagonal $\Lambda = \text{diag}(R_{ii}/|R_{ii}|)$
    \item Compute $A' = Q\Lambda$, where $A'$ is a Haar-random unitary
\end{enumerate}
We then let the qiskit \cite{qiskit} compiler find native gates that best approximates this unitary $A'$.
\begin{figure}[htb]
    \centering
    \includegraphics[width=0.49\textwidth]{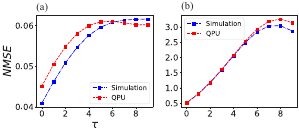}
    \caption{\textbf{QPU Test:} NMSE for various $\tau$ using \textbf{(a)} Ising chain and \textbf{(b)} Mackey-Glass sequences.}
    \label{fig:QPU_results}
\end{figure}

In Fig.~\ref{fig:QPU_results}, we compare the NMSE of our QPU experiment with that of the simulation, both run with $10^4$ shots on the Ising chain and Mackey-Glass sequence. We note that both the simulation and the QPU runs have inherently limited predictive capabilities (as seen in the very high NMSE values) because of the greatly reduced training and test lengths. Nonetheless, our main objective here is to demonstrate that the proposed method can be executed on current quantum hardware.

\subsection{Echo state property analysis}
We now present the results of the experiment on the analysis of the echo state property of different reservoir computing models. In all methods, we replicate the reservoirs as described in the respective references. In particular:
\begin{itemize}
    \item For the classical minimal ESN model \cite{Lukoeviius2012} -- we use a network of the form: 
    \begin{equation} x_{t+1} = (1-\alpha)x_t + \alpha \tanh(W_{in}s_t + b + Wx_t) \end{equation} where \(x_k\) is 1000-dimensional. Each element in \(W_{\mathrm{in}}, b,\) and \(W\) is initialized with the random variable \(X \sim \mathcal{N}(-0.5, 1)\). Next, \(W\) is renormalized so that its spectral radius is 1.25, and \(\alpha\) is set to 0.3. This configuration reproduces a model that performed well on the Mackey--Glass sequence used throughout this paper. Each run was performed with a different initialization of \(x_0\), with each component randomly drawn from a uniform distribution between 0 and 1.
    \item For the feedback-driven QRC protocol \cite{feedbackdriven} -- we take the architecture proposed in Fig.~\ref{fig:All_QRC_Regimes} (c)  with the hyperparameters $a_{in} = 0.001$, $a_{fb} = 2.5$ and a Haar-randomly chosen $U_{res}$, using a state vector simulator to compute the expectation values of the measurements. Each run was performed with a different Haar-random initialization of the start state.
    \item For the QRC scheme with a mid-circuit measurement architecture \cite{mid_circuit_measurements} -- we take the architecture proposed in Fig.~\ref{fig:All_QRC_Regimes} (d) with 10 time steps and setting $a=5$ and $10^4$ shots\footnote{Although no specific value for $a$ was explicitly given in the original work, we were able to reproduce similar results to original work's NARMA2 benchmark results when setting $a=5$.}. Each run was performed with a different Haar-random 2-qubit initial state while keeping the 2 ancilla qubits always starting in the $\ket{0}$ state.   
    \item For the model proposed in this paper -- we use the model described in Fig.~\ref{fig:All_QRC_Regimes} (e) with $a_{in} = 1$, $a_{fb} = 1.6$, a Haar-randomly chosen $U_{\text{Haar}}$ and $10^4$ shots. Each run was performed with a different Haar-random 2-qubit initial state (denoted by $\rho_0$ in Fig.~\ref{fig:All_QRC_Regimes} (e)).
\end{itemize}

The results of the ESP analysis are shown in Fig.~\ref{fig:Convergence}. We see that all models aside from the proposed model with post-measurement resets, clearly exhibit the echo-state property. And even for the proposed model with post-measurement resets, we see that effect of the initial state becomes less prominent over time, although it does not completely vanish. This shows that all the QRC models investigated exhibit ESPs to at least some extent. To make sure that we did not stumble upon a special case, we found qualitatively similar results when we repeating the analysis with several different initializations of the fixed random parameters (e.g., $W_{in}$, $U_{\text{Haar}}$, etc.) and over different dimensions of the internal state of the reservoir.
During the course of our experiments there was no post-selection of the results, and all results from the different circuits composed of the various uniquely randomly generated $U_{Haar}$ matrices are included in the statistics .

\section{Conclusion}
In this paper, we propose a quantum reservoir computing scheme that uses mid-circuit measurements and feedback within the runtime of a qubit. Our initial experiments aim to determine whether feedback connection that happen during the coherence time of the qubit actually improves memory capacity, rather than merely introducing uncertainties due to measurement noise. We show that such an architecture with built-in resets has non-zero memory capacity for past inputs, indicating that feedback connections have an impact. In particular, peaks in memory capacity at specific feedback strengths are consistent with improved performance on prediction tasks involving both classical and quantum chaotic systems. 
Our ESP results show that most investigated QRC regimes exhibit the ESP, while our proposed model with post-measurement resets seems to have a mild form of the ESP.

 This work demonstrates sufficiently encouraging results to warrant further investigations into the approach of having feedback connections during coherence time. Given that our model exhibits some initial memory and prediction capabilities even at very small scales, exploring the scaling of our model is a very promising direction for future investigation that can reveal more on the effectiveness of this approach. Further investigations could also include experimenting with alternative reset strategies, such as using conditional resets or omitting resets altogether, thus allowing the model to continue to evolve from the post-measurement state while also having feedback connections.

\appendices
\section{Noise Test}
\label{sec:noise-test}
\begin{figure}[htb]
    \centering
    \includegraphics[width=0.80\linewidth]{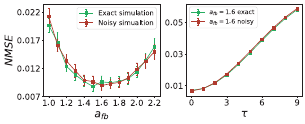}
    \caption{\textbf{Noise Test:} We analyze how noise in our QRC regime affects the NMSE in predicting the dynamics of the Ising chain.}
    \label{fig:noisee}
\end{figure}

We present preliminary noise tests for larger runs, accounting for the significant reductions in QPU runs. We use the same run statistics as in our numerical experiments. In our shot-based simulation, we introduce a depolarization (DP) channel after each unitary of the form:
\begin{equation}
    DP(\rho) = (1-\lambda) \rho + \lambda \frac{I}{d}
\end{equation}
where $\lambda$ is the probability of the state to be found in the completely mixed state $\frac{I}{d}$. For the one qubit case, the DP channel can expressed also with the Kraus operators: \begin{align}&K_{0}=\sqrt{1-\frac{3\lambda}{4}}I, K_1 = \sqrt{1-\frac{\lambda}{4}}X,\\&K_2 = \sqrt{1-\frac{\lambda}{4}}Y, K_3 = \sqrt{1-\frac{\lambda}{4}}Z\end{align}
We set lambda to $0.04$, which corresponds geometrically to a uniform Bloch sphere reduction of $4\%$, or in terms of random processes: With $97\%$ chance we apply the identity. With $1\%$ chance we use $X$, $Y$ or $Z$. The predictive capabilities shown in Fig.~\ref{fig:noisee} on the Ising chain (similar to Fig.~\ref{fig:results} (b)) demonstrate that our model is robust to some amount of depolarizing noise.
\section{Code Availability}
The Python implementation can be found at \url{https://github.com/Muri7910/FB_Con_MCM_QRC}.


\bibliographystyle{IEEEtran}
\bibliography{IEEEabrv,mybib.bib}

\begin{thebibliography}{10}
\providecommand{\url}[1]{#1}
\csname url@samestyle\endcsname
\providecommand{\newblock}{\relax}
\providecommand{\bibinfo}[2]{#2}
\providecommand{\BIBentrySTDinterwordspacing}{\spaceskip=0pt\relax}
\providecommand{\BIBentryALTinterwordstretchfactor}{4}
\providecommand{\BIBentryALTinterwordspacing}{\spaceskip=\fontdimen2\font plus
\BIBentryALTinterwordstretchfactor\fontdimen3\font minus \fontdimen4\font\relax}
\providecommand{\BIBforeignlanguage}[2]{{%
\expandafter\ifx\csname l@#1\endcsname\relax
\typeout{** WARNING: IEEEtran.bst: No hyphenation pattern has been}%
\typeout{** loaded for the language `#1'. Using the pattern for}%
\typeout{** the default language instead.}%
\else
\language=\csname l@#1\endcsname
\fi
#2}}
\providecommand{\BIBdecl}{\relax}
\BIBdecl

\bibitem{best_practice}
C.~Wringe, M.~Trefzer, and S.~Stepney, ``Reservoir computing benchmarks: a tutorial review and critique,'' \emph{arXiv [cs.ET]}, 2024.

\bibitem{class_reser}
M.~Dale, S.~O'Keefe, A.~Sebald, S.~Stepney, and M.~A. Trefzer, ``\BIBforeignlanguage{en}{Reservoir computing quality: connectivity and topology},'' \emph{\BIBforeignlanguage{en}{Nat. Comput.}}, vol.~20, no.~2, pp. 205--216, Jun. 2021.

\bibitem{class_reser_2}
\BIBentryALTinterwordspacing
X.-Y. Duan, X.~Ying, S.-Y. Leng, J.~Kurths, W.~Lin, and H.-F. Ma, ``Embedding theory of reservoir computing and reducing reservoir network using time delays,'' \emph{Phys. Rev. Res.}, vol.~5, p. L022041, May 2023. [Online]. Available: \url{https://link.aps.org/doi/10.1103/PhysRevResearch.5.L022041}
\BIBentrySTDinterwordspacing

\bibitem{class_reser_3}
M.~Dale, J.~F. Miller, S.~Stepney, and M.~A. Trefzer, ``\BIBforeignlanguage{en}{A substrate-independent framework to characterize reservoir computers},'' \emph{\BIBforeignlanguage{en}{Proc. Math. Phys. Eng. Sci.}}, vol. 475, no. 2226, p. 20180723, Jun. 2019.

\bibitem{quant_reser_1}
\BIBentryALTinterwordspacing
K.~Fujii and K.~Nakajima, ``Harnessing disordered-ensemble quantum dynamics for machine learning,'' \emph{Phys. Rev. Appl.}, vol.~8, p. 024030, Aug 2017. [Online]. Available: \url{https://link.aps.org/doi/10.1103/PhysRevApplied.8.024030}
\BIBentrySTDinterwordspacing

\bibitem{quant_reser_2}
A.~Senanian, S.~Prabhu, V.~Kremenetski, S.~Roy, Y.~Cao, J.~Kline, T.~Onodera, L.~G. Wright, X.~Wu, V.~Fatemi, and P.~L. McMahon, ``\BIBforeignlanguage{en}{Microwave signal processing using an analog quantum reservoir computer},'' \emph{\BIBforeignlanguage{en}{Nat. Commun.}}, vol.~15, no.~1, p. 7490, Aug. 2024.

\bibitem{quant_reser_3}
D.~Fry, A.~Deshmukh, S.~Y.-C. Chen, V.~Rastunkov, and V.~Markov, ``\BIBforeignlanguage{en}{Optimizing quantum noise-induced reservoir computing for nonlinear and chaotic time series prediction},'' \emph{\BIBforeignlanguage{en}{Sci. Rep.}}, vol.~13, no.~1, p. 19326, Nov. 2023.

\bibitem{quant_reserv_4}
J.~Dudas, B.~Carles, E.~Plouet, F.~A. Mizrahi, J.~Grollier, and D.~Markovi{\'c}, ``\BIBforeignlanguage{en}{Quantum reservoir computing implementation on coherently coupled quantum oscillators},'' \emph{\BIBforeignlanguage{en}{Npj Quantum Inf.}}, vol.~9, no.~1, Jul. 2023.

\bibitem{feedbackdriven}
K.~Kobayashi, K.~Fujii, and N.~Yamamoto, ``\BIBforeignlanguage{en}{Feedback-driven quantum reservoir computing for time-series analysis},'' \emph{\BIBforeignlanguage{en}{PRX quantum}}, vol.~5, no.~4, Nov. 2024.

\bibitem{weak_meas}
P.~Mujal, R.~Mart{\'\i}nez-Pe{\~n}a, G.~L. Giorgi, M.~C. Soriano, and R.~Zambrini, ``\BIBforeignlanguage{en}{Time-series quantum reservoir computing with weak and projective measurements},'' \emph{\BIBforeignlanguage{en}{Npj Quantum Inf.}}, vol.~9, no.~1, Feb. 2023.

\bibitem{rewind_protocol}
\BIBentryALTinterwordspacing
J.~Chen, H.~I. Nurdin, and N.~Yamamoto, ``Temporal information processing on noisy quantum computers,'' \emph{Phys. Rev. Appl.}, vol.~14, p. 024065, Aug 2020. [Online]. Available: \url{https://link.aps.org/doi/10.1103/PhysRevApplied.14.024065}
\BIBentrySTDinterwordspacing

\bibitem{quantum_dyn}
K.~Fujii and K.~Nakajima, ``\BIBforeignlanguage{en}{Harnessing disordered-ensemble quantum dynamics for machine learning},'' \emph{\BIBforeignlanguage{en}{Phys. Rev. Appl.}}, vol.~8, no.~2, Aug. 2017.

\bibitem{boosting}
K.~Nakajima, K.~Fujii, M.~Negoro, K.~Mitarai, and M.~Kitagawa, ``\BIBforeignlanguage{en}{Boosting computational power through spatial multiplexing in quantum reservoir computing},'' \emph{\BIBforeignlanguage{en}{Phys. Rev. Appl.}}, vol.~11, no.~3, Mar. 2019.

\bibitem{Maass2002-wv}
W.~Maass, T.~Natschl{\"a}ger, and H.~Markram, ``\BIBforeignlanguage{en}{Real-time computing without stable states: a new framework for neural computation based on perturbations},'' \emph{\BIBforeignlanguage{en}{Neural Comput.}}, vol.~14, no.~11, pp. 2531--2560, Nov. 2002.

\bibitem{mid_circuit_measurements}
\BIBentryALTinterwordspacing
T.~Yasuda, Y.~Suzuki, T.~Kubota, K.~Nakajima, Q.~Gao, W.~Zhang, S.~Shimono, H.~I. Nurdin, and N.~Yamamoto, ``Quantum reservoir computing with repeated measurements on superconducting devices,'' 2023. [Online]. Available: \url{https://arxiv.org/abs/2310.06706}
\BIBentrySTDinterwordspacing

\bibitem{https://doi.org/10.48550/arxiv.2503.17939}
\BIBentryALTinterwordspacing
T.~Monomi, W.~Setoyama, and Y.~Hasegawa, ``Feedback-enhanced quantum reservoir computing with weak measurements,'' 2025. [Online]. Available: \url{https://arxiv.org/abs/2503.17939}
\BIBentrySTDinterwordspacing

\bibitem{dyn_circ_1}
\BIBentryALTinterwordspacing
E.~B\"aumer, V.~Tripathi, D.~S. Wang, P.~Rall, E.~H. Chen, S.~Majumder, A.~Seif, and Z.~K. Minev, ``Efficient long-range entanglement using dynamic circuits,'' \emph{PRX Quantum}, vol.~5, p. 030339, Aug 2024. [Online]. Available: \url{https://link.aps.org/doi/10.1103/PRXQuantum.5.030339}
\BIBentrySTDinterwordspacing

\bibitem{Macke-Glass}
M.~C. Mackey and L.~Glass, ``\BIBforeignlanguage{en}{Oscillation and chaos in physiological control systems},'' \emph{\BIBforeignlanguage{en}{Science}}, vol. 197, no. 4300, pp. 287--289, Jul. 1977.

\bibitem{naturalQRC}
Y.~Suzuki, Q.~Gao, K.~C. Pradel, K.~Yasuoka, and N.~Yamamoto, ``\BIBforeignlanguage{en}{Natural quantum reservoir computing for temporal information processing},'' \emph{\BIBforeignlanguage{en}{Sci. Rep.}}, vol.~12, no.~1, p. 1353, Jan. 2022.

\bibitem{moore1920reciprocal}
E.~H. Moore, ``On the reciprocal of the general algebraic matrix,'' \emph{Bulletin of the american mathematical society}, vol.~26, pp. 294--295, 1920.

\bibitem{penrose1955generalized}
R.~Penrose, ``A generalized inverse for matrices,'' in \emph{Mathematical proceedings of the Cambridge philosophical society}, vol.~51, no.~3.\hskip 1em plus 0.5em minus 0.4em\relax Cambridge University Press, 1955, pp. 406--413.

\bibitem{stm}
\BIBentryALTinterwordspacing
Z.~Liao, H.~Yamahara, K.~Terao, K.~Ma, M.~Seki, and H.~Tabata, ``Short-term memory capacity analysis of lu3fe4co0.5si0.5o12-based spin cluster glass towards reservoir computing,'' \emph{Scientific Reports}, vol.~13, no.~1, p. 5260, 2023. [Online]. Available: \url{https://doi.org/10.1038/s41598-023-32084-8}
\BIBentrySTDinterwordspacing

\bibitem{ESN-Mackey}
\BIBentryALTinterwordspacing
Y.~Li, K.~Hu, K.~Nakajima, and Y.~Pan, ``Composite force learning of chaotic echo state networks for time-series prediction,'' in \emph{2022 41st Chinese Control Conference (CCC)}.\hskip 1em plus 0.5em minus 0.4em\relax IEEE, Jul. 2022, p. 7355–7360. [Online]. Available: \url{http://dx.doi.org/10.23919/CCC55666.2022.9901897}
\BIBentrySTDinterwordspacing

\bibitem{ESN_Mackey_2}
L.~Appeltant, M.~C. Soriano, G.~Van~der Sande, J.~Danckaert, S.~Massar, J.~Dambre, B.~Schrauwen, C.~R. Mirasso, and I.~Fischer, ``\BIBforeignlanguage{en}{Information processing using a single dynamical node as complex system},'' \emph{\BIBforeignlanguage{en}{Nat. Commun.}}, vol.~2, no.~1, p. 468, Sep. 2011.

\bibitem{PhysRevLett.64.2215}
\BIBentryALTinterwordspacing
H.-J. St\"ockmann and J.~Stein, ````quantum'' chaos in billiards studied by microwave absorption,'' \emph{Phys. Rev. Lett.}, vol.~64, pp. 2215--2218, May 1990. [Online]. Available: \url{https://link.aps.org/doi/10.1103/PhysRevLett.64.2215}
\BIBentrySTDinterwordspacing

\bibitem{chaotic}
\BIBentryALTinterwordspacing
M.~C. Ba\~nuls, J.~I. Cirac, and M.~B. Hastings, ``Strong and weak thermalization of infinite nonintegrable quantum systems,'' \emph{Phys. Rev. Lett.}, vol. 106, p. 050405, Feb 2011. [Online]. Available: \url{https://link.aps.org/doi/10.1103/PhysRevLett.106.050405}
\BIBentrySTDinterwordspacing

\bibitem{jaeger2001echo}
H.~Jaeger, ``The “echo state” approach to analysing and training recurrent neural networks-with an erratum note,'' \emph{Bonn, Germany: German national research center for information technology gmd technical report}, vol. 148, no.~34, p.~13, 2001.

\bibitem{PhysRevApplied.8.024030}
\BIBentryALTinterwordspacing
K.~Fujii and K.~Nakajima, ``Harnessing disordered-ensemble quantum dynamics for machine learning,'' \emph{Phys. Rev. Appl.}, vol.~8, p. 024030, Aug 2017. [Online]. Available: \url{https://link.aps.org/doi/10.1103/PhysRevApplied.8.024030}
\BIBentrySTDinterwordspacing

\bibitem{Chen2019-nt}
J.~Chen and H.~I. Nurdin, ``\BIBforeignlanguage{en}{Learning nonlinear input--output maps with dissipative quantum systems},'' \emph{\BIBforeignlanguage{en}{Quantum Inf. Process.}}, vol.~18, no.~7, Jul. 2019.

\bibitem{PhysRevE.110.024207}
\BIBentryALTinterwordspacing
S.~Kobayashi, Q.~H. Tran, and K.~Nakajima, ``Extending echo state property for quantum reservoir computing,'' \emph{Phys. Rev. E}, vol. 110, p. 024207, Aug 2024. [Online]. Available: \url{https://link.aps.org/doi/10.1103/PhysRevE.110.024207}
\BIBentrySTDinterwordspacing

\bibitem{Lukoeviius2012}
\BIBentryALTinterwordspacing
M.~Lukoševičius, \emph{A Practical Guide to Applying Echo State Networks}.\hskip 1em plus 0.5em minus 0.4em\relax Springer Berlin Heidelberg, 2012, p. 659–686. [Online]. Available: \url{http://dx.doi.org/10.1007/978-3-642-35289-8_36}
\BIBentrySTDinterwordspacing

\bibitem{Sannia2024}
\BIBentryALTinterwordspacing
A.~Sannia, R.~Martínez-Peña, M.~C. Soriano, G.~L. Giorgi, and R.~Zambrini, ``Dissipation as a resource for quantum reservoir computing,'' \emph{Quantum}, vol.~8, p. 1291, Mar. 2024. [Online]. Available: \url{http://dx.doi.org/10.22331/q-2024-03-20-1291}
\BIBentrySTDinterwordspacing

\bibitem{OliviaDiMatteo2021}
O.~D. Matteo, ``Understanding the haar measure,'' \url{https://pennylane.ai/qml/demos/tutorial_haar_measure}, 03 2021, date Accessed: 2025-04-09.

\bibitem{Haar_random}
F.~Mezzadri, ``\BIBforeignlanguage{English}{How to generate random matrices from the classical compact groups},'' \emph{\BIBforeignlanguage{English}{Notices of the American Mathematical Society}}, vol.~54, no.~5, pp. 592 -- 604, May 2007.

\bibitem{qiskit}
\BIBentryALTinterwordspacing
A.~Javadi-Abhari, M.~Treinish, K.~Krsulich, C.~J. Wood, J.~Lishman, J.~Gacon, S.~Martiel, P.~D. Nation, L.~S. Bishop, A.~W. Cross, B.~R. Johnson, and J.~M. Gambetta, ``Quantum computing with qiskit,'' 2024. [Online]. Available: \url{https://arxiv.org/abs/2405.08810}
\BIBentrySTDinterwordspacing

\end{thebibliography}

\end{document}